\documentclass[aps,twocolumn,showpacs]{revtex4}

\usepackage{graphicx}
\usepackage{dcolumn}
\usepackage{bm}

\begin{document}

\title{Critical wave-packet dynamics in the power-law bond disordered Anderson Model}

\author{R.\ P.\ A.\  Lima, F.\ A.\ B.\ F. de Moura, and M.\ L.\ Lyra}

\affiliation{ Departamento de F\'{\i}sica, Universidade Federal de Alagoas,
Macei\'{o} AL 57072-970, Brazil}

\author{H.\ N.\ Nazareno}

\affiliation{ICCMP, Universidade de Bras\'{\i}lia, CP 04513, 70919-970,
Bras\'{\i}lia-DF,Brazil }

\begin{abstract}

We investigate the wave-packet dynamics of the power-law bond
disordered one-dimensional Anderson model with hopping amplitudes
decreasing as $H_{nm}\propto |n-m|^{-\alpha}$. We consider the
critical case ($\alpha=1$). Using an exact diagonalization scheme
on finite chains, we compute the participation moments of all
stationary energy eigenstates as well as the spreading of an
initially localized wave-packet. The eigenstates multifractality
is characterized by the set of fractal dimensions of the
participation moments. The wave-packet shows a diffusive-like
spread developing a power-law tail and achieves a stationary
non-uniform profile after reflecting at the chain boundaries. As a
consequence, the time-dependent participation moments exhibit two
distinct scaling regimes. We formulate a finite-size scaling
hypothesis for the participation moments relating their scaling
exponents to the ones governing the return probability and
wave-function power-law decays.
\end{abstract} \maketitle

\section{Introduction}

Non-interacting electron systems with uniformly distributed
disorder usually show an Anderson transition from localized to
extended states. In general, such transition occurs only for
spatial dimensions greater than $d=2$ in the case of systems with
short-range hopping,  a result supported by a single parameter
scaling theory~\cite{anderson,abrahams}. On the other hand, when
long-range couplings are assumed, a transition from localized to
delocalized electronic states can be found even in 1D disordered
systems\cite{prbm1,prbm2}. In this case, one has an interplay between
the hopping range and the degree of disorder. The former favors
propagation while the later inhibits it. It is worthwhile to
mention that propagation of carriers was also obtained in
low-dimensional models with short-range hopping but presenting
correlated disorder, such as random dimer chains
\cite{flores1,dunlap,dimer,bellani} and in chains with scale-free
disorder\cite{chico,izrailev,mauricio2d}, as well as in
chains containing quasi-periodic structures, as for example Fibonacci,
Thue-Morse, Harper sequences ~\cite{brito,harper,ryu}.

On the other hand, for an ordered 1D system with  hopping terms
decaying with a power-law with exponent $\alpha$, it was shown
that for $\alpha>2$ one recovers the result for the
tight-binding model~\cite{Nazareno99}. More interesting is the
behavior corresponding to $0<\alpha<1$. For
$\alpha=0$ an initially localized wave-packet presents {\it self
trapping}, i.e., the particle performs oscillations in a definite
region of the lattice, visiting periodically the starting
position. By increasing $\alpha$ the localization is lost. When
the power exponent equals unity, and for sufficient short times,
the packet diffuses with a diffusion coefficient that increases
with the number of sites. This effect is absent in the model with
only nearest-neighbor hopping~\cite{Nazareno99}.

More recently, it was studied the dynamics of an electron in a
one-dimensional Anderson model with non-random hoppings falling
off as some power $\alpha$ of the distance between
sites~\cite{brito2}. It was found that the larger the hopping
range, the more extended the wave-packet becomes as time evolves.
When the disorder is increased, the wave-packet tends to remain
more localized. For a low degree of disorder, the exponent
$\alpha=1.5$ indicates the onset for fast propagation. Moreover,
the inclusion of a dc electric field introduces the effect of
dynamical localization. The fast propagation found for
$\alpha<1.5$ is in agreement with the reported delocalization of
 states located close to one of the band edges~\cite{adame00,adame03,adame04b}.

The power-law random band matrix (PRBM) model also exhibits a
delocalization transition~\cite{prbm1,prbm2}. This model describes
one-dimensional electronic systems with random long-range hopping
amplitudes with standard deviation decaying as $1/r^{\alpha}$ for
sites at a distance $r>>b$, where $b$ is a typical bandwidth. It
was shown that at $\alpha=1$ it presents an Anderson-like
transition with all states being localized for $\alpha>1$ and
extended for $\alpha<1$.  At the critical point $\alpha=1$, the
inverse participation ratio distribution, the wave-functions
multifractal spectra, the level statistics and the time-evolution
of the wave-packet size have been investigated both analytically
and
numerically~\cite{prbm1,prbm2,lima3,izrailevdi1,izrailevdi2,tessieri}.
Within the same spirit of the PRBM, a model for non-interacting
electrons in a $2D$ lattice with random on-site potentials and
random power-law decaying transfer terms was numerically
investigated by exploring the finite-size scaling properties of
the fluctuations in the mean level spacing\cite{potempa}. It was
found that the one-electron eigenstates become extended for
transfer terms decaying slower than $1/r^2$.  Finally, the
Anderson transition in a $1D$ chain with random power-law decaying
hopping terms and non-random on-site energies was numerically characterized in ref.~\cite{lima4}.

The moments of the position and those of the related probability
density are known to exhibit different scaling
behaviors~\cite{luck}. This feature reflects the complexity of the
scaling laws governing the dynamics of quantum
systems~\cite{geisel}. In this work, we investigate the critical
dynamics of the power-law bond disordered Anderson model
($\alpha=1$) by looking at the moments of the wave-function
probability distribution. Using an exact diagonalization scheme on
finite chains, we compute the participation moments of all 
energy eigenstates and follow the time-evolution of an initially 
localized wave packet. After that, we perform an average of the participation
moments over different configurations and energies, which allow us
to compute the critical exponents $D_q$ associated with the 
multifractal character of the critical eigenfunctions. Examining the time-evolution of an initially localized
wave-packet, we obtain the decay exponent of the autocorrelation
function ($C(t) \propto t^{-D_2}$) and the size dependence of the
asymptotic return probability ($R(t)\propto N^{-D_2^{\Psi}}$). We
will employ a finite-size scaling analysis of the time dependent
participation number moments, relating their scaling exponents
with the power-law exponents of the evolving wave-packet.
\begin{figure}[t!]
\centerline{\includegraphics[width=70mm,clip]{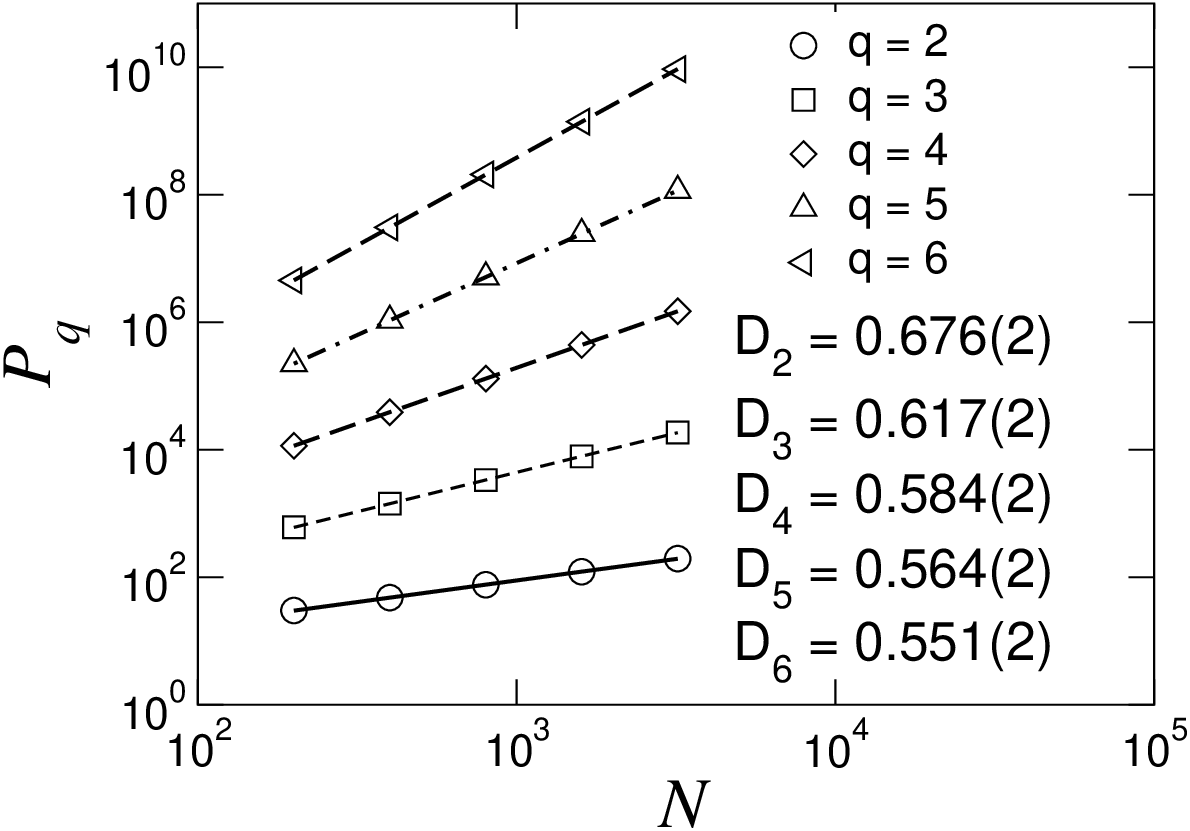}}
\caption{ The average participation moment ($P_q$) as a function
of the chain size ($N$) for several values of $q$ ranging from $2$
to $6$. The distinct fractal dimensions ($P_q\propto N^{D_q(q-1)})$
exhibited by the participation moments reflect the multifractality
of the critical eigenstates.} \label{fig1}
\end{figure}

\section{Model, Formalism and Results}

\subsection{Model Hamiltonian}

We consider a single electron in a 1D chain with open
boundaries, described by the Anderson Hamiltonian
\begin{equation}
H = \sum _{n\neq m}^{N} H_{nm}|n\rangle\langle m| ~~~,
\end{equation}
where $|n\rangle$ represents the state with the electron localized
at site $n$. In the present pure random bond Anderson model, the
on-site potentials $\epsilon_i$ are site independent and in Eq.~1
were taken to be $\epsilon_i=0$ without any loss of generality.
Long-range disorder is introduced by assuming the hopping amplitudes $H_{ij}$
to be randomly distributed and also displaying a  power-law decay.
We will, hereafter, consider
\begin{equation}
H_{nm} = W_{nm}/|n-m|^{\alpha}~~~,
\end{equation}
where $W_{nm}$ is a random variable with a uniform distribution in
the interval $[-1,+1]$. As a function of the exponent $\alpha$
characterizing the decay of the hopping amplitudes, this model
displays a localization-delocalization transition at $\alpha =
1$~\cite{lima4} in close connection with the
PRBM~\cite{prbm1,prbm2}. For off-diagonal terms
decaying slower than $1/|n-m|$, i.e., for $\alpha<1$, all states
become delocalized. At $\alpha=1$ the states are critical and the
level spacing statistics is between the Poison and Wigner
surmises. In the limit of $\alpha\rightarrow\infty$ one recovers
the 1D Anderson model with just first-neighbors random hopping
amplitudes.

\subsection{Eigen-functions participation moments}

We will be particularly interested in investigating the
participation moments  and the time-evolution of an initially
localized wave-packet  at the critical point  $\alpha =1$. The
participation moments for a particular disorder configuration and
eigenstate will be defined as the inverse of the moments of the
probability density
\begin{equation}
P^j_q=\frac{1}{\sum_{n=1}^N|f^j_n|^{2q}}
\end{equation}
where $f_n^{(j)}$ is the  amplitude at site $n$ of the $j$-th
eigenstate  obtained from an exact diagonalization scheme on
finite chains with sizes ranging from $N=400$ up to $N=3200$
sites. Here we used $32\times 10^3$ states for each chain size  in
order to average over distinct disorder configurations and
energies
\begin{figure}[t!]
\centerline{\includegraphics[width=70mm,clip]{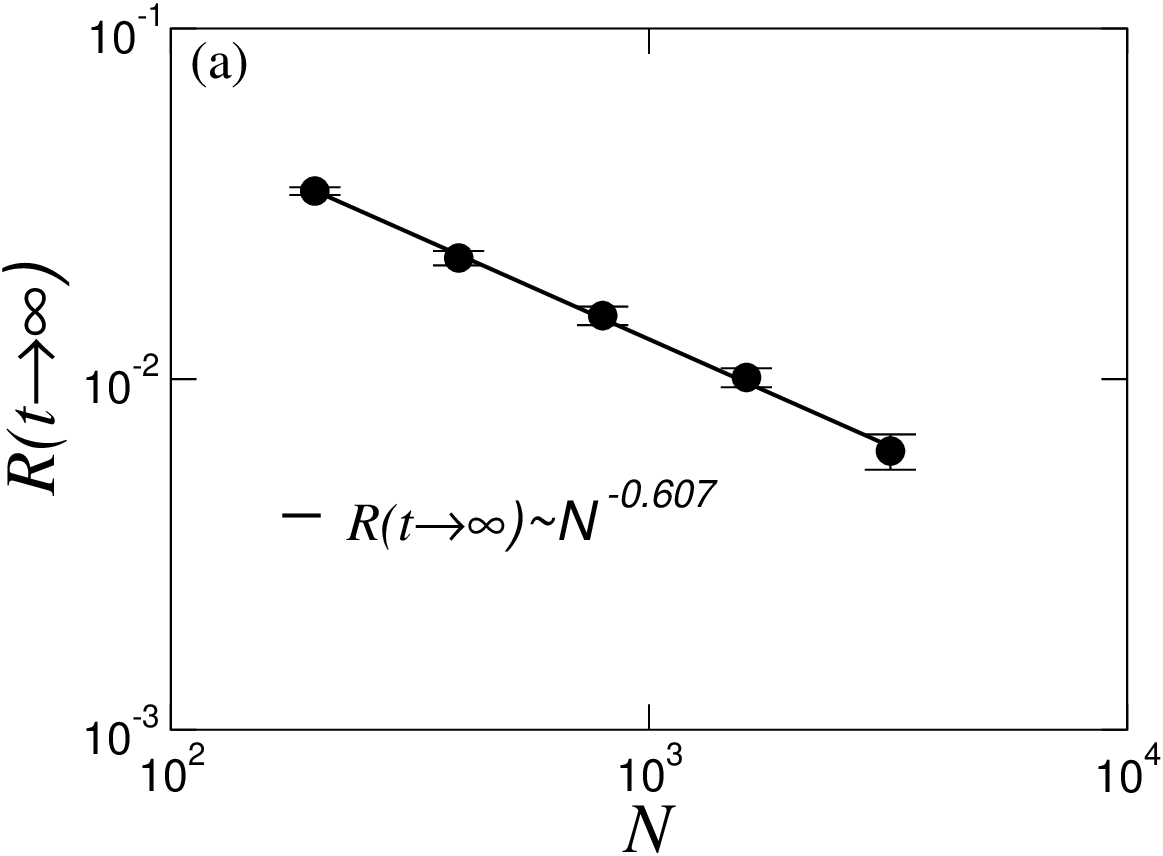}}
\centerline{\includegraphics[width=70mm,clip]{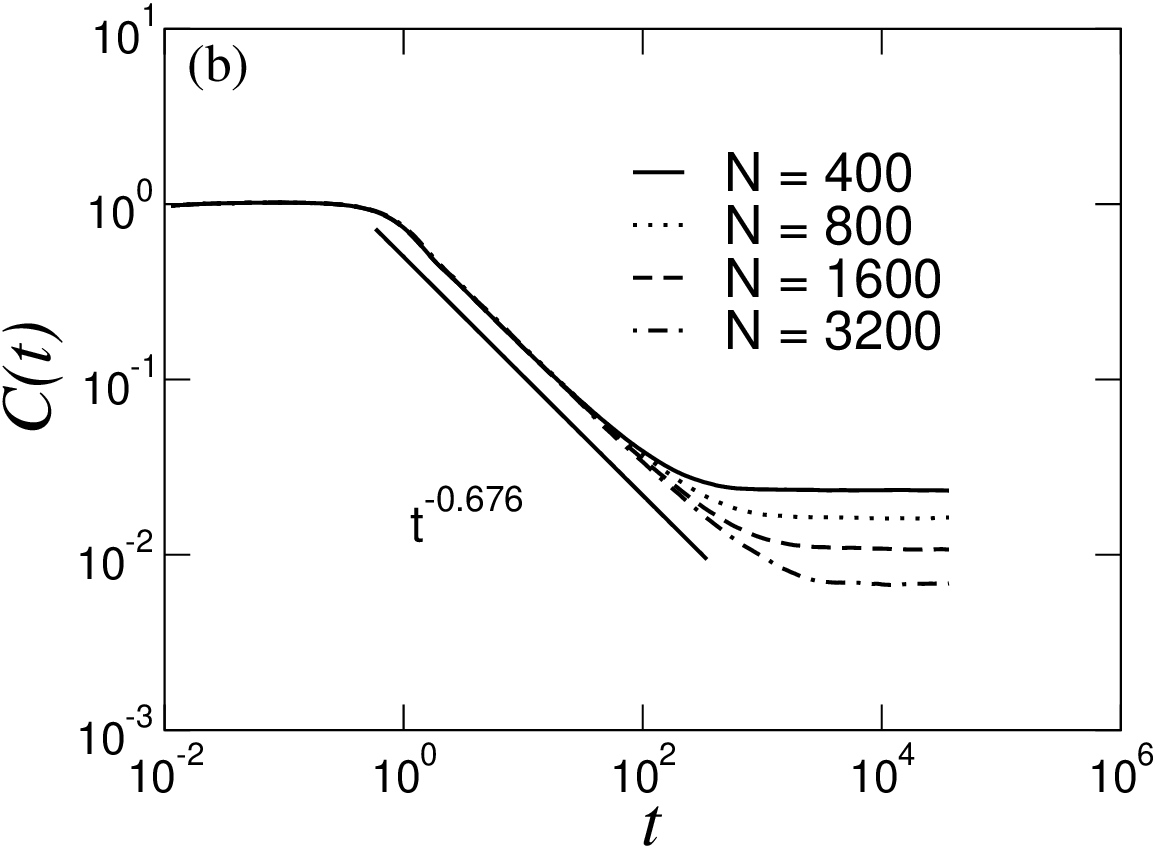}}
\caption{(a)The return probability function for $t \rightarrow
\infty$ as a function of  the number of sites. The small power-law decay
exponent $R(t\rightarrow\infty)\propto N^{-D_2^{\Psi}}$, with
$D_2^{\Psi}=0.607(10)$, is consistent with a wave-function with a
slowly decaying envelope. (b) Autocorrelation function versus time
$t$ for chains with $N=400$ up to $N=3200$ sites. The power-law
decay for intermediate times with an exponent $D_2=0.676$ is in
good agreement with the previous numerical estimate for $D_2$.}
\label{fig2}
\end{figure}
\begin{equation}
 P_q= \left\langle\frac{\sum_{j=1}^N
P^j_q}{N} \right\rangle, 
\end{equation}
 where $< >$ stands for
configuration average. Therefore, $P_q$ probes the average $q$-th
moment of the critical stationary states. In fig.~1 we show the
finite-size scaling of different moments ranging from $q=2$ up to
$q=6$. As it has been shown by Wegner\cite{wegner}, all moments
show a power-law dependence on the form ($P_q \propto
N^{D_q(q-1)}$). Each moment has its characteristic exponent $D_q$
as indicated in Fig.~ (1) which reflects the multifractality of
the critical eigenstates. Our results indicate that the fractal dimension $D_q$ slowly
decreases from $D_2=0.676$ towards the asymptotic value
$D_{\infty}=0.5$.

\subsection{Wave-packet Dynamics}

In order to obtain  the time-evolution of an initially localized
wave-packet ($|\Phi(t=0)\rangle$), we expand the wave-function in
the Wannier representation
\begin{equation}
|\Phi(t)\rangle=\sum_n f_n(t)|n\rangle.
\end{equation}
 We   solve the time-dependent
Schr\"odinger equation for the wave-function components $f_n(t)$
($\hbar=1$)
\begin{equation}
i\frac{df_n(t)}{dt}=\sum _{n\neq
m}^{N}\frac{W_{nm}}{|n-m|^{\alpha}}f_n(t),
\end{equation}
using the numerical formalism proposed in ref.~\cite{naza2003}:
\begin{equation}
|\Phi(t)\rangle={\bf U}^{\dagger}\exp{(-i{\bf D}t)}{\bf U}|\Phi(t=0)\rangle,
\end{equation}
where ${\bf D}$ is the  diagonal form of the Hamiltonian and ${\bf U}$ is a unitary  matrix.
\begin{figure}[t!]
\centerline{\includegraphics[width=70mm,clip]{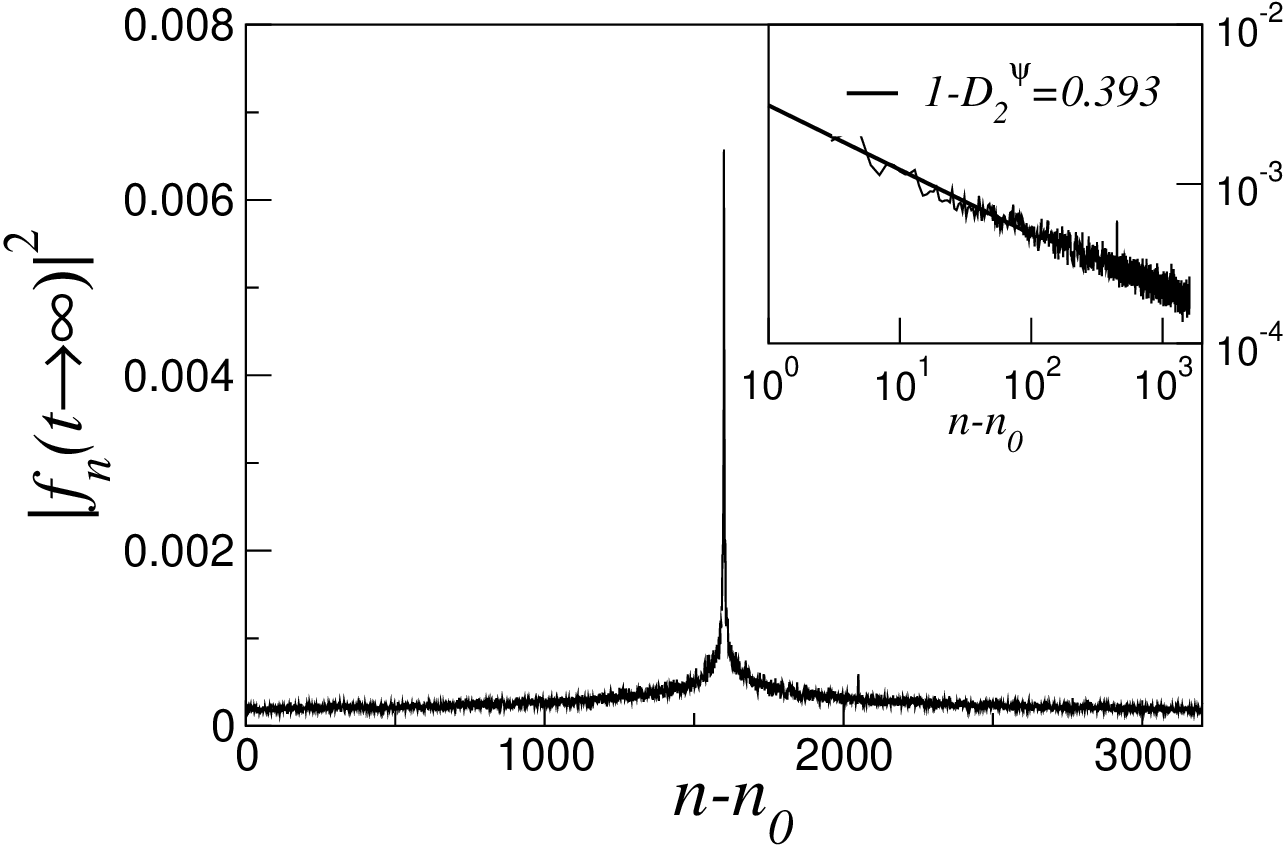}}
\caption{Average wave-function profile after a very long evolution
time ($t \rightarrow \infty$). In the inset, we note that the
envelope displays a power-law decay with an exponent
$1-D_2^{\Psi}=0.393$, is agreement with the measure exponent of
the asymptotic return probability. } \label{fig3}
\end{figure}
In what follows we consider the electron initially localized in a single site, i.e. $|\Phi(t=0)\rangle=|n_0\rangle$. We start by discussing the behavior of the autocorrelation
function:
\begin{equation}
 C(t)=\frac{1}{t}\int_0^{t}{R(t)dt},
\end{equation}
where $R(t)=|f_{n_0}(t)|^2$  denotes the return probability. The
time dependent autocorrelation function provides information about
localization of the wave-packet as well as of its fractal
character. In fact, after an initial waiting time,  the
autocorrelation function shall behave as:
\begin{equation}
C(t)\propto t^{-D_2},
\end{equation}
for one-dimensional systems, which be definition is the same
power-law decay presented by the return probability. In fig.~2(a)
we plot the return probability for $t \rightarrow \infty$ as a
function of the number of sites. After the initial decay due to the
wave-function spread, it saturates at a size-dependent plateau
which decreases as $1/N^{D_2^{\Psi}}$. In fig.~2(b), the
autocorrelation function versus time $t$ is shown for different
chain sizes. It exhibits a power dependence for intermediate times
whose exponent is in good agreement with the previous estimate for
$D_2$. Its saturation for $t\rightarrow\infty$ has the same
finite-size origin presented by the return probability.

The asymptotic return probability decaying slower than $1/N$
indicates that the statistically stationary wave-function does not
assume a uniform profile. In fig.~(3) we plot the average
stationary wave-function profile achieved long after the
wave-function reflection at the chain boundaries. It develops a
power-law decaying tail with
$|f_n(t \rightarrow \infty)|^2 \propto 1/(n-n_0)^{1-D_2^{\Psi }}$
(see inset), as expected according
to the wave-function normalization constraint. Before reaching the
chain boundaries, the average wave-function envelope can be well
described by
\begin{equation}
|f(n,t)|^2 = R(t)(n-n_0+1)^{-(1-D_2^{\Psi})}
\end{equation}
for $n-n_0<N^*(t)$, where $N^*(t)$ is a time dependent cutoff
after which the wave-function decays exponentially and, therefore, 
can be considered as having a vanishing amplitude. The return probability decays as $R(t)\propto
t^{-D_2}$. It is straightforward to show, using the normalization
criterium, that the cutoff shall scale as $N^*(t)\propto
t^{D_2/D_2^{\Psi}}$. The above asymptotic power-law shape of the
wave-function is in agreement with the previous prediction for the
wave-packet spreading in quantum systems with fractal energy
spectra and eigenfunctions\cite{geisel}.
\begin{figure}[t!]
\centerline{\includegraphics[width=70mm,clip]{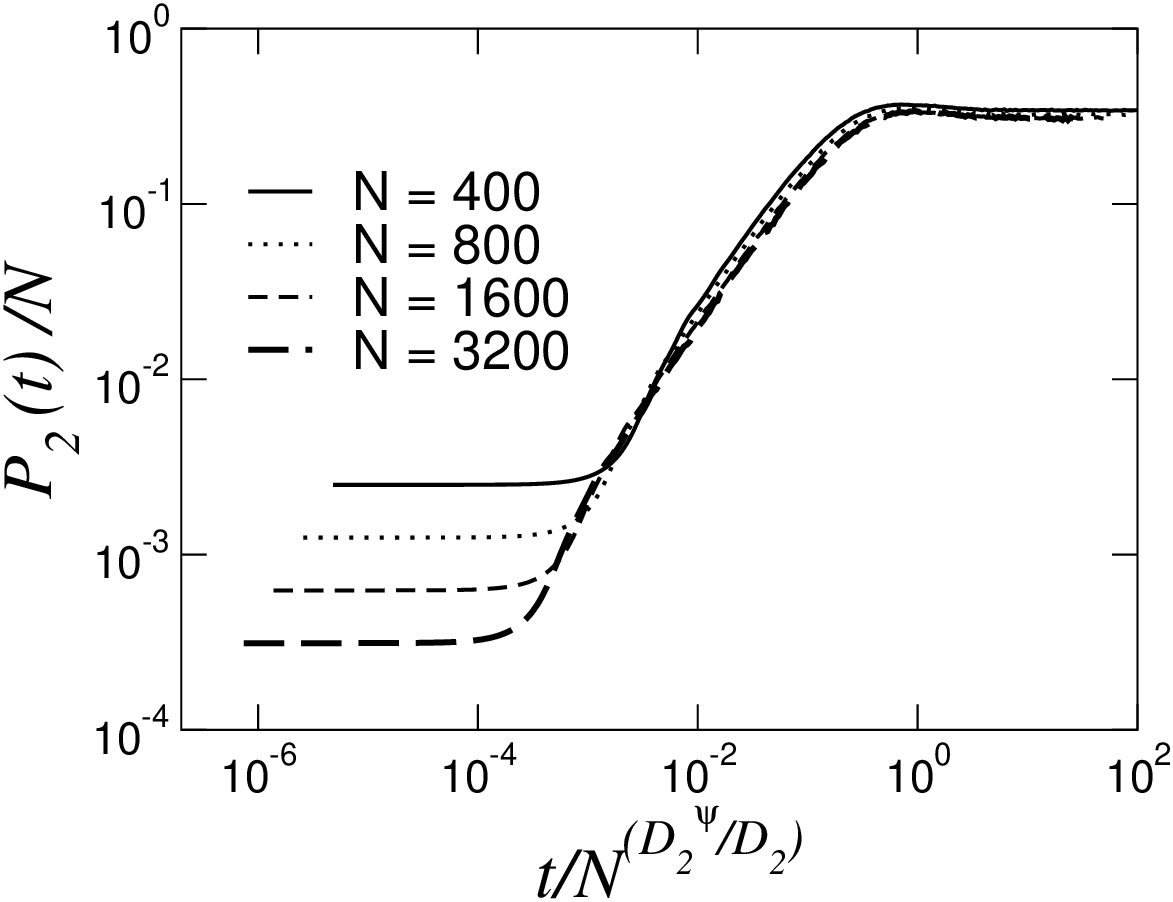}}
\caption{The normalized average participation number $P_2/N$ as a
function of the scaled time ($t/N^{D_2^{\Psi}/D_2}$).  After a
short transient time, the data collapse indicates that the
participation number grows as $t^{D_2/D_2^{\Psi}}$ and saturates
after the wave-function reflection at the chain boundaries. The
destructive interference which sets up during the reflection at
the boundaries is signaled by a maximum in $P_2$. The data
collapse corroborates the scaling hypothesis in Eq.~(12). }
\label{fig4}
\end{figure}
\begin{figure}[t!]
\centerline{\includegraphics[width=70mm,clip]{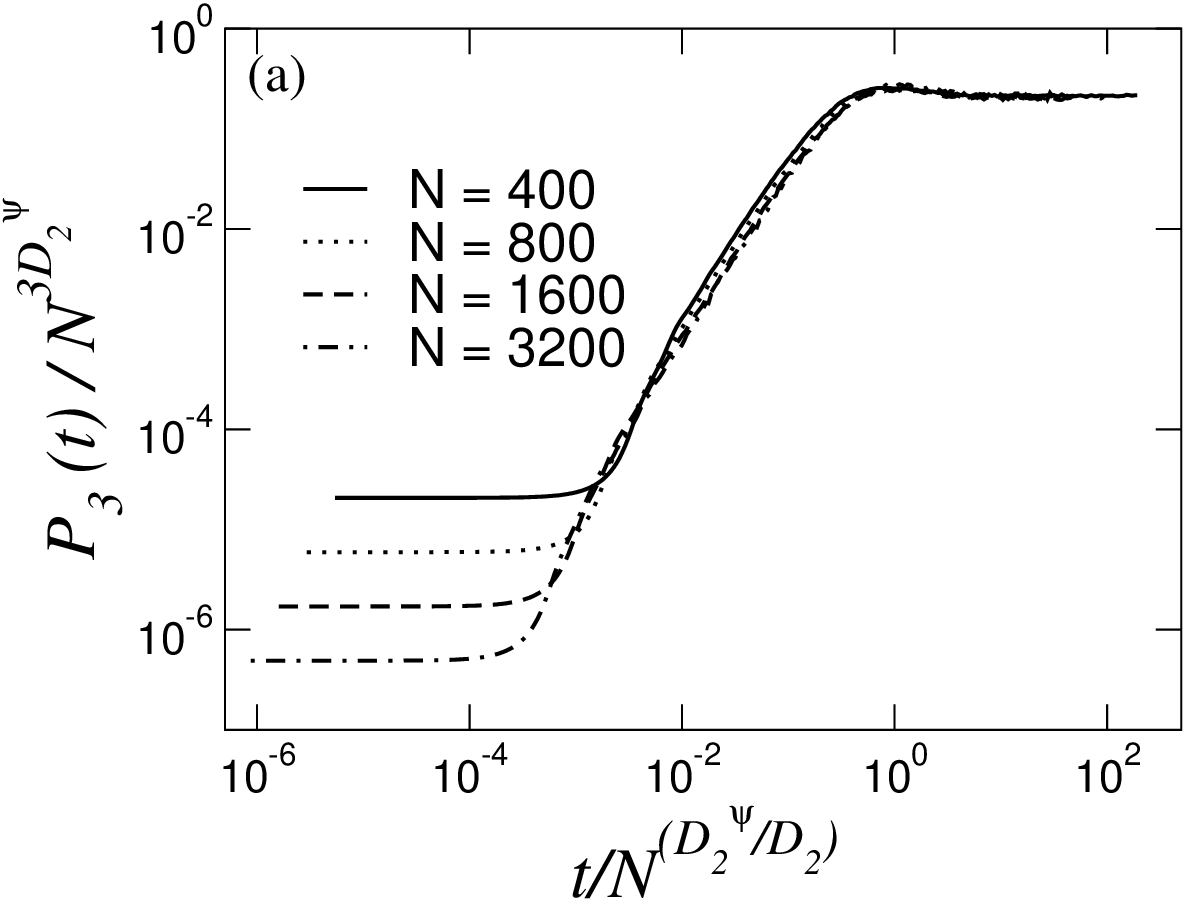}}
\centerline{\includegraphics[width=70mm,clip]{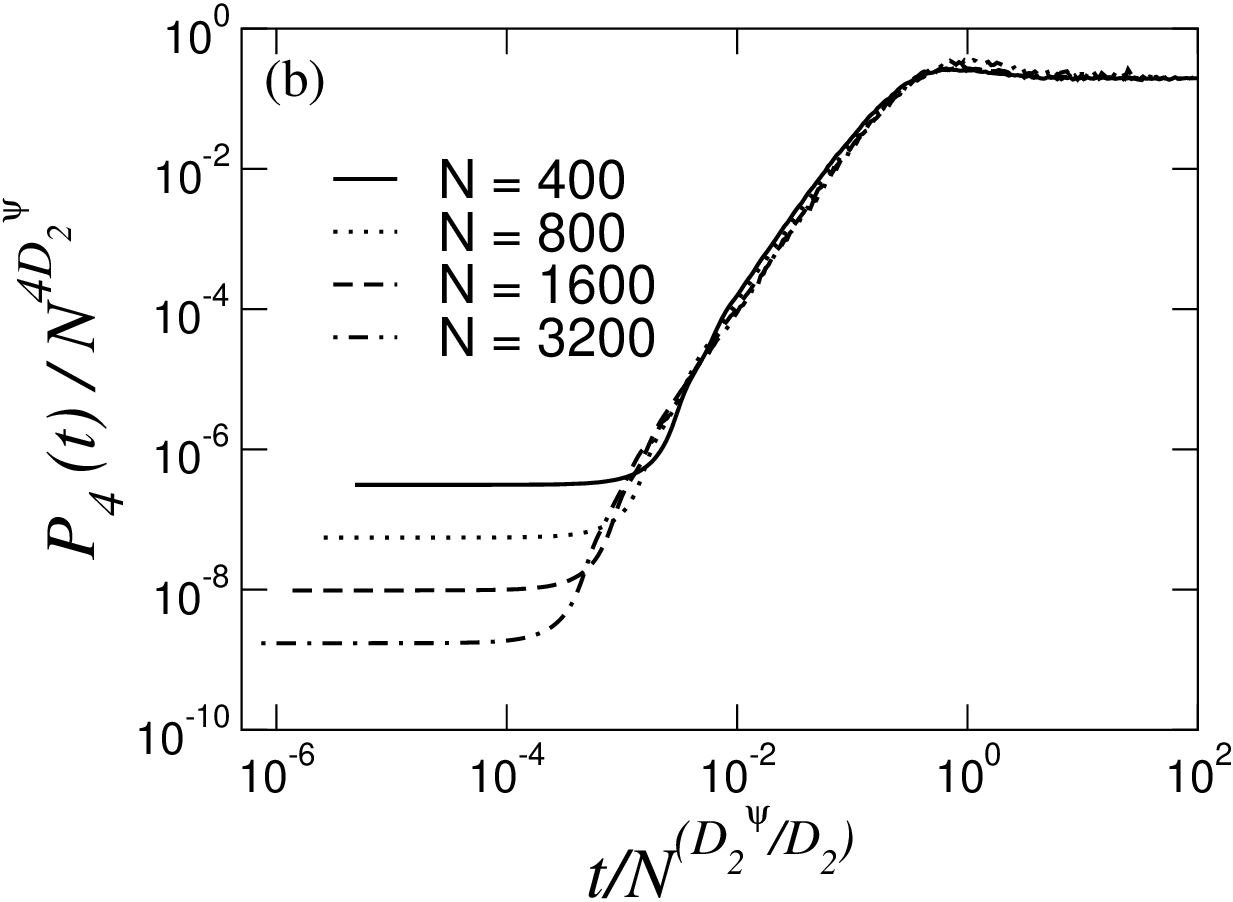}}
\centerline{\includegraphics[width=70mm,clip]{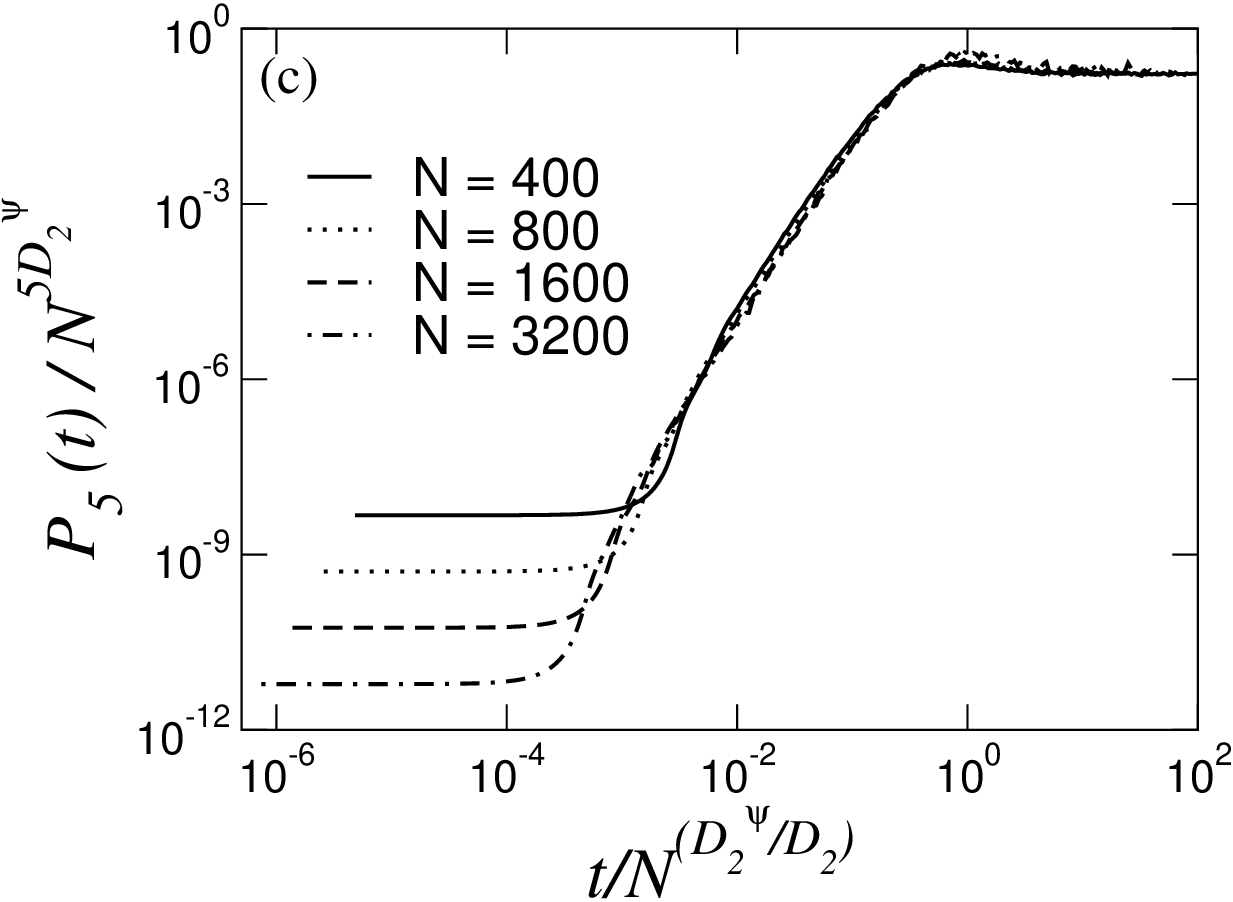}}
\caption{Scaled participation  moments $P_q(N,t)/N^{qD_2^{\Psi}}$
as a function of the scaled time ($t/N^{D_2^{\Psi}/D_2}$)  for
$q=3,4,5$. The data collapse corroborates
 the scaling hypothesis in Eq.~(15). The maximum, also appearing
 in fig.~4, signals the destructive interference between the incident
 and reflected waves near the chain boundaries.} \label{fig5}
\end{figure}
We  further calculated the time dependent participation moments
\begin{equation}
P_q(t)= \left\langle \frac{1}{\sum_n^N|f_n(t)|^{2q}}\right\rangle.
\end{equation}
The second moment ($P_2$) has been commonly used as a measurement
of the number of sites that participate in the wave function. For
a wave-packet fully localized at a single site, $P_2=1$, while for
a uniform distribution of the wave function,  $P_2$  reaches the
maximum value $N$. In general, $P_2$ is size independent in the
localized phase and scales linearly with the system size for
extended states.

In the present case, for which the wave-function develops the power-law
form given by Eq.(10), the participation moments present a
non-trivial scaling. Two distinct behaviors can be anticipated.
For $q<1/(1-D_2^{\Psi})$, the participation moment shall display a
power-law growth (after a short waiting time) on the form
$P_q(t)\propto t^{(q-1)D_2/D_2^{\Psi}}$, followed by a saturation
after reaching the chain boundaries. The saturation plateau is
predicted to scale as $P(N,t\rightarrow\infty)\propto N^{(q-1)}$.
Both time regimes can be represented by a single finite-size
scaling hypothesis for the participation moment which can be
written as
\begin{equation}
P_q(t,N)=N^{(q-1)}g(t/N^{D_2^{\Psi}/D_2})
\end{equation}
with $g(\infty)=cte$ and $g(x\rightarrow 0)\propto
x^{(q-1)D_2/D_2^{\Psi}}$. In fig.~4, we show our numerical results
for the scaled second moment ($P_2/N$) versus scaled time
($t/N^{D_2^{\Psi}/D_2}$), which falls in the above regime of
$q<1/(1-D_2^{\Psi})$. These data were  averaged over $100$ time
histories considering distinct disorder configurations. The second
moment remains constant during a very short initial waiting time
before the spread effectively starts to take place. After that, we
found that the above scaling hypothesis reproduces very well the
participation time-evolution as data from distinct chain sizes
could be collapsed in the same curve. As anticipated, the
participation moment has a power-law growth followed by a
saturation. The slight deviation from perfect collapse  is due to
corrections to scaling which are more pronounced for the smaller
system sizes. The maximum displayed by the participation number
reflects the destructive interference between the incident and
reflected waves near the chain boundaries.

For $q>1/(1-D_2^{\Psi})$, the presence of a cutoff does not
influence the initial spread of the participation moments.
Therefore, the initial growth is completely determined by the
return probability decay and results in $P_q(t)\propto t^{qD_2}$.
On the other hand, the scaling of the asymptotic plateau only
depends on the wave-function decay exponent, being given by
$P(N,t\rightarrow\infty)\propto N^{qD_2^{\Psi}}$. For this regime,
the proper scaling form of the participation moment reads
\begin{equation}
P_q(t,N)=N^{qD_2^{\Psi}}g'(t/N^{D_2^{\Psi}/D_2})
\end{equation}
with $g'(\infty)=cte$ and $g'(x\rightarrow 0)\propto x^{qD_2}$.
 In figure~5 we show the scaling
analysis of participation moments for this regime ($q=3,4,$ and $5$).
The trends are mainly the same as that presented by $P_2$, namely
data collapse with saturation for long times and a power-law
growth for intermediate times preceded by  a size independent
initial waiting time. However, the participation moment scaling
exponent now depends on the wave-function decay exponent as
predicted by Eq.(13).

In both regimes, the characteristic time scale after that
finite-size effects starts to take place scales as $t^*\propto
N^{D_2^{\Psi}/D_2}$, with $D_2^{\Psi}/D_2=0.90(2)$. The fact that
this scaling exponent is smaller than unity (and therefore corresponds to a faster 
than ballistic dynamics) is related to the non-local nature of the
long-range couplings. Finally, for exponentially localized
wave-functions $[P_q]^{1/(q-1)}$ is proportional to the
localization length for any $q$. Using $N^*$ as a measure of the
characteristic size of the present power-law decaying wave-packet,
this feature is only true for $q<1/(1-D_2^{\Psi})$. For  higher
moments it displays a sub-linear size dependence whose exponent
continuously decreases towards $D_2^{\Psi}$ as
$q\rightarrow\infty$.

\section{conclusions}

We investigated the  one-dimensional Anderson model with
off-diagonal disorder and matrix elements $H_{ij}$ decaying as
$1/|i-j|^{\alpha}$ for $\alpha=1$ (critical regime). Using an
exact  diagonalization scheme on finite chains, we computed the
participation moments of all energy eigenstates and reported the
critical exponents $D_q$ associated with the multifractal
character of the stationary states.  Examining the time-evolution
of an initially localized wave-packet, we observed that the
wave-function develops a power-law tail in the form
$|\Psi(n-n_o)|^2 \propto (n-n_0)^{-(1-D_2^{\Psi})}$ with
$D_2^{\Psi}=0.607(10)$. It achieves a stationary non-uniform
profile after reflecting at the chain boundaries. As a
consequence, the time-dependent participation moments $P_q(t)$
exhibit a non-trivial finite-size scaling. For
$q<1/(1-D_2^{\Psi})$, the participation moments grow in time as
$P_q\propto t^{(q-1)D_2/D_2^{\Psi}}$, where the exponent
$D_2=0.676(2)$ governs the decay of the return probability as well
as the one of the autocorrelation function. It saturates in a
plateau proportional to $N^{q-1}$. Therefore, the usual
participation number $P_2$ reaches a value proportional to the
chain size as occurs with a uniformly distributed state. Higher
order participation moments with $q>1/(1-D_2^{\Psi})$ grow in time
as $t^{qD_2}$ and saturate at a plateau proportional to
$N^{qD_2^{\Psi}}$. We used a finite-size scaling hypothesis to put
the participation moments in a universal form as a function of the
reduced variable $t/N^{D_2^{\Psi}/D_2}$. As such, the
characteristic size of the wave-packet grows in time as
$N^*\propto t^{D_2/D_2^{\Psi}}$, which for the present model with
long-range couplings results in a faster than ballistic spread.
The proposed dynamic scaling relations are expected to remain
valid for general model systems of waves spreading in a power-law
fashion over a random medium\cite{geisel}. The special character
of the present model, with all energy eigenstates being
multifractal, provides a clear picture of the complex dynamic
scaling which takes place at the vicinity of the Anderson
transition.

\section{acknowledgements}

This work was partially supported by the Brazilian research
agencies CNPq (Conselho Nacional de Pesquisa) and  by the Alagoas State research agency FAPEAL
(Funda\c{c}\~ao de Amparo a Pesquisa do Estado de Alagoas).

\noindent

\end{document}